\documentclass[11pt]{article}

\usepackage{graphicx}
\usepackage{amsmath}
\usepackage{hyperref}
\usepackage{url}
\usepackage{natbib}

\usepackage[margin=1in]{geometry}

\graphicspath{{./}}

\begin{document}

\title{TREMORS: An Agentic Assistant for Multi-Datacenter Seismic Data Acquisition}

\author{
Ryley G. Hill$^{1,*}$ \\
\texttt{rghill@lanl.gov} \\
\and
Richard Alfaro-Diaz$^{1}$ \\
\and
Jonathan MacCarthy$^{1}$ \\
\and
Jonas A. Kintner$^{1}$ \\
\and
Christopher W. Johnson$^{1}$ \\
\\
\small $^{1}$EES-17 National Security Earth Science, Los Alamos National Laboratory, Los Alamos, NM, USA \\
\small $^{*}$Corresponding author
}

\date{}

\maketitle

\begin{abstract}
Seismology increasingly depends upon large data retrieval across multi-datacenter platforms.
Yet, data procurement often demands domain expertise, user burden, and is difficult to reproduce.
As archives continue to grow, translating scientific intent into structured workflows that operate across multiple repositories and produce high quality, AI ready data is becoming an increasingly urgent challenge.
We present TREMORS (Text Referenced Event Mapping and Output Renderer for Seismographs), an agentic framework that uses large language model reasoning within a constrained execution graph to automate seismic data retrieval.
TREMORS translates natural language queries into a structured intermediate schema, which drives execution through a constrained LangGraph workflow.
Example workflows demonstrate support for both event-based and continuous waveform acquisition.
The framework is designed to extend across heterogeneous multi-datacenter systems through a portable schema and modular workflow components.
This work positions agentic workflows as the catalyst that will connect scientific intent with distributed seismic data systems, enabling a future of reproducible, data-driven inquiry while reducing the burden of routine acquisition tasks.
\end{abstract}

\section{Introduction}

Observational seismology requires reliable access to global coverage of seismic network waveforms and metadata that are distributed across many datacenters \citep{ahern2003fdsn}.
Datacenters are unique in that each institution controls the archival method and service interfaces for seismic data \citep{davis2024knowledge}, while providing open access repositories.
Community efforts have standardized access to station metadata, event catalogs, and waveform time series \citep{ahern2003fdsn,beyreuther2010obspy}.
These advancements have made global and regional seismic observations more accessible enabling global-scale data-driven studies that encompass large tectonic environments, long timescales, and multiple networks.

The task of finding, retrieving, and integrating data remains a barrier in some scientific workflows, specifically in procurement.
Observational seismology increasingly requires combining information from multiple datacenters, reconciling heterogeneous metadata and station conventions, correctly combining overlapping event catalogs, and constructing waveform requests that vary in duration \citep{krischer2015obspy}.
For experts, these hurdles are often routine, but rarely are they trivial.
They demand familiarity with service and metadata conventions as well as an understanding of station and network naming structures \citep{newman2013wilber}.
Idiosyncrasies of individual providers can further exacerbate data retrieval barriers, especially for newcomers of the field.
Data procurement is frequently time consuming, difficult to reproduce, and unnecessarily dependent on specialist knowledge.
The challenge does not solely affect new users, but exists as a rudimentary time sink for experienced researchers who often rebuild similar workflows for different projects.

Large language models (LLMs) and LLM-agent enabled workflows, or agentic models, create an opportunity to revise the seismic network data procurement pipeline.
In multiple scientific disciplines workflows utilizing LLM-agents are accelerating and automating considerable portions of routine tasks \citep{xi2025rise,gottweis2025towards,casper2025ai}.
Agentic workflows are composed of autonomous artificial intelligence (AI) agents that integrate reasoning and tool-based execution to autonomously plan and complete complex, multi-step tasks with minimal human intervention \citep{gottweis2025towards,casper2025ai}.
The important advancement is not merely natural language interaction, but the introduction of machine-actionable workflow layers that can reason over user intent, choose among data sources, enforce valid query structure, and document each decision in a reproducible format.
These models combine semantic interpretation with constrained execution.
Essentially, from a user request the agent will map the natural language query onto a structured intermediate representation, then use that representation to drive downstream actions while reasoning about its current state.
The approach preserves many benefits of automation while retaining reproducibility required for scientific use.

LLM-agents have broader implications than convenience alone.
Tools that mediate between scientific intent and distributed seismic infrastructure will alter how researchers interact with data archives and how acquisition pipelines across projects are standardized.
Notably, they provide a path toward an AI-ready semiological infrastructure in which retrieval will be organized around portable schemas and modular workflows rather than any single access protocol.
Under that conceptual model, current web-service standards are valuable implementation targets \citep{newman2013wilber,pickle2026seed,fdsn_webservices}, but not the ideal endpoint.
The long-term goal is a flexible acquisition framework that can merge results across multiple repositories, easily scale as new APIs and metadata services are developed, and adapt to cloud-based repositories as archives continue to increase in data volume.

Current services provide a useful foundation for this transition.
In particular, the International Federation of Digital Seismograph Networks (FDSN) web services have demonstrated the value of common protocols for accessing station metadata and waveform data across distributed archives \citep{ahern2003fdsn,fdsn_services,fdsn_webservices}.
These services are essential for modern seismological practice.
However, in the context of next-generation workflows, they are but a single example of standardized access.
The more general challenge is how to build retrieval systems that operate across multiple repositories, reconcile heterogeneous responses, and remain adaptable as standards evolve.

Here, we present TREMORS (Text Referenced Event Mapping \& Output Renderer for Seismographs), an agentic seismology assistant for automated waveform and metadata procurement.
TREMORS translates natural language requests into a structured query schema and executes them through a constrained workflow of data acquisition.
While the present implementation is demonstrated using FDSN compatible services, the wider contribution is a workflow pattern for multiple datacenter access that we intentionally leave portable for future API integration.
TREMORS is intended not only as a practical tool, but as an example of how agentic systems will contribute in observational seismology science and move toward more reproducible, accessible, and accelerated data-driven scientific workflows.

\section{Methods}
\subsection{Agentic Architecture}

The TREMORS agentic architecture is conceptually based on the Universal Research and Scientific Agent (URSA) framework developed at Los Alamos National Laboratory \citep{grosskopf2025ursa}, but supports its own agentic backend architecture.
Multiple architectures have emerged that combine LLMs with planning, memory, and tool-use capabilities \citep{casper2025ai}.
Since there is no established definition of \emph{AI agent} we give careful consideration to the term \emph{agent}.
An agent enables multi-step reasoning, dynamic decision making, and vary substantially in terms of autonomy, reproducibility, security constraints, and domain specializations.
What is advantageous about the URSA framework is that the LLM operates as a reasoner within a structured LangChain/LangGraph \citep{langchain_website} workflow and the reasoning determines how predefined execution nodes are invoked rather than calling arbitrary tools.
This constrained node-based design enables goal-driven planning while preserving reproducibility, safety, and auditability in the scientific pipeline.
The structured graph we implement does not require a sophisticated LLM to call and works well with many lightweight open models that a user can download locally.
In this application we use gpt-oss:120b  that contains 120 billion parameters and requires $\sim$65GB of disk storage.
In the TREMORS repository are instructions for setting up Ollama to download open LLMs \cite{HillZenodoTREMORS}.

The agentic architecture preserves many benefits of automation while retaining reproducibility required for scientific use.
Specifically the architecture is designed with built in subject-matter-expert workflows that allows the model to use this structured process as guardrails for the LLM.
For data access the TREMORS agent requires a backend and interfaces with FDSN webservices through the ObsPy API \citep{beyreuther2010obspy}.
This design choice does not limit the agent to a single access point and is adaptable to additional backends.
The ObsPy API provides integration across multiple datacenters, e.g., EARTHSCOPE, ORFEUS, GEOFON, SCEDC, NCEDC, etc., for routine automation.
The URSA framework is applied to this API through user prompting to identify earthquakes based on magnitude, location, and time and constructs targeted fdsnws-dataselect queries for corresponding waveform data.

\subsection{Prompt Engineering for Deterministic Agent Routing}
User prompting must enforce a well-defined structure that constrains the output to ensure reliable execution within the LangGraph agentic workflow.
Structured prompts guide the model toward predictable, machine-interpretable outputs.
Vague LLM prompts introduce ambiguity and produce incorrect branching decisions, yet, surprisingly still perform reasonable well with inputs as "Give me big earthquakes".
In TREMORS the LLM is instructed to translate user queries into a JSON object with a fixed schema representing seismic query parameters, e.g., temporal bounds, spatial constraints, magnitude thresholds, and waveform retrieval options.
This schema acts as an intermediate representation between natural language and downstream tools.
The key design principles are:
\begin{itemize}
    \item \textbf{Schema Enforcement:} The prompt explicitly defines all allowable fields and their types, e.g., float, string, or boolean, ensuring consistent output formatting.
    \item \textbf{Default Behavior:} Sensible defaults reduce underspecification in user queries, e.g., datacenter selection, time windows, and preprocessing parameters.
    \item \textbf{Mutual Exclusivity and Conditional Logic:} Certain fields are constrained through explicit instructions to prevent invalid combinations, e.g., continuous waveform requests or event-based queries.
    \item \textbf{Semantic Parsing to Parameters:} Natural language expressions are mapped to quantitative fields, e.g., ``near'' to radius, ``large earthquakes'' to magnitude thresholds, and ``last decade'' to date ranges.
    \item \textbf{Execution-Aware Flags:} Boolean fields directly control downstream graph nodes, enabling deterministic routing through the workflow, e.g., \texttt{get\_waveforms} or \texttt{plot\_waveforms}.
\end{itemize}

By enforcing this structured interface, the language model functions as a semantic parser whose output can be deterministically consumed by the agent graph. This significantly reduces failure modes associated with ambiguous intent and ensures that each query follows a valid execution path through the system. The objects that the parser deduces include:

\begin{enumerate}
    \item \texttt{datacenter}: string (e.g., ``ISC'', ``USGS'', ``IRIS''), inferred or explicit. Default: ``ISC''
    \item \texttt{date$_{min}$}: UTCISO string (e.g., 2010-01-01T00:00:00). Default: 1970-01-01T00:00:00
    \item \texttt{date$_{max}$}: UTCISO string
    \item \texttt{lat$_{min}$}: float
    \item \texttt{lat$_{max}$}: float
    \item \texttt{lon$_{min}$}: float
    \item \texttt{lon$_{max}$}: float
    \item \texttt{depth$_{min}$}: float (minimum depth in km)
    \item \texttt{depth$_{max}$}: float (maximum depth in km)
    \item \texttt{mag$_{min}$}: float (minimum magnitude)
    \item \texttt{mag$_{max}$}: float (maximum magnitude)
    \item \texttt{get\_waveforms}: boolean (true if user requests downloading waveforms for discovered events)
    \item \texttt{plot\_waveforms}: boolean (true if user requests plotting waveforms)
    \item \texttt{get\_continuous\_waveforms}: boolean (true only if user explicitly requests continuous waveform data without event-based filtering)
    \item \texttt{stations\_file}: string (path to a \texttt{test\_stations.txt} file if provided)
    \item \texttt{net}: string (network filter, e.g., ``CI'', ``IU'')
    \item \texttt{sta}: string (station filter, e.g., ``ANMO'')
    \item \texttt{loc}: string (location filter, e.g., ``00'', ``--'')
    \item \texttt{chan}: string (channel filter, e.g., ``BHZ'')
    \item \texttt{parallel}: integer (number of threads for download). Default: 4
    \item \texttt{bulk\_chunk}: integer (number of requests per bulk call). Default: 50
    \item \texttt{dir\_date}: boolean (true if user specifies organizing output by date or YYYY/DOY). Default: false
    \item \texttt{dir\_stat}: boolean (true if user specifies organizing output by station or network/station). Default: false
    \item \texttt{response}: boolean (true if user requests response or XML metadata download). Default: false
    \item \texttt{radius}: float (search radius if user specifies ``near'' or ``within'')
    \item \texttt{radius\_unit}: string (unit of radius, e.g., ``km'' or ``deg'')
    \item \texttt{pre\_event\_sec}: float (seconds before event for waveform window). Default: 30.0
    \item \texttt{post\_event\_sec}: float (seconds after event for waveform window). Default: 600.0
\end{enumerate}

\begin{figure}
    \centering
    \includegraphics[width=0.65\linewidth]{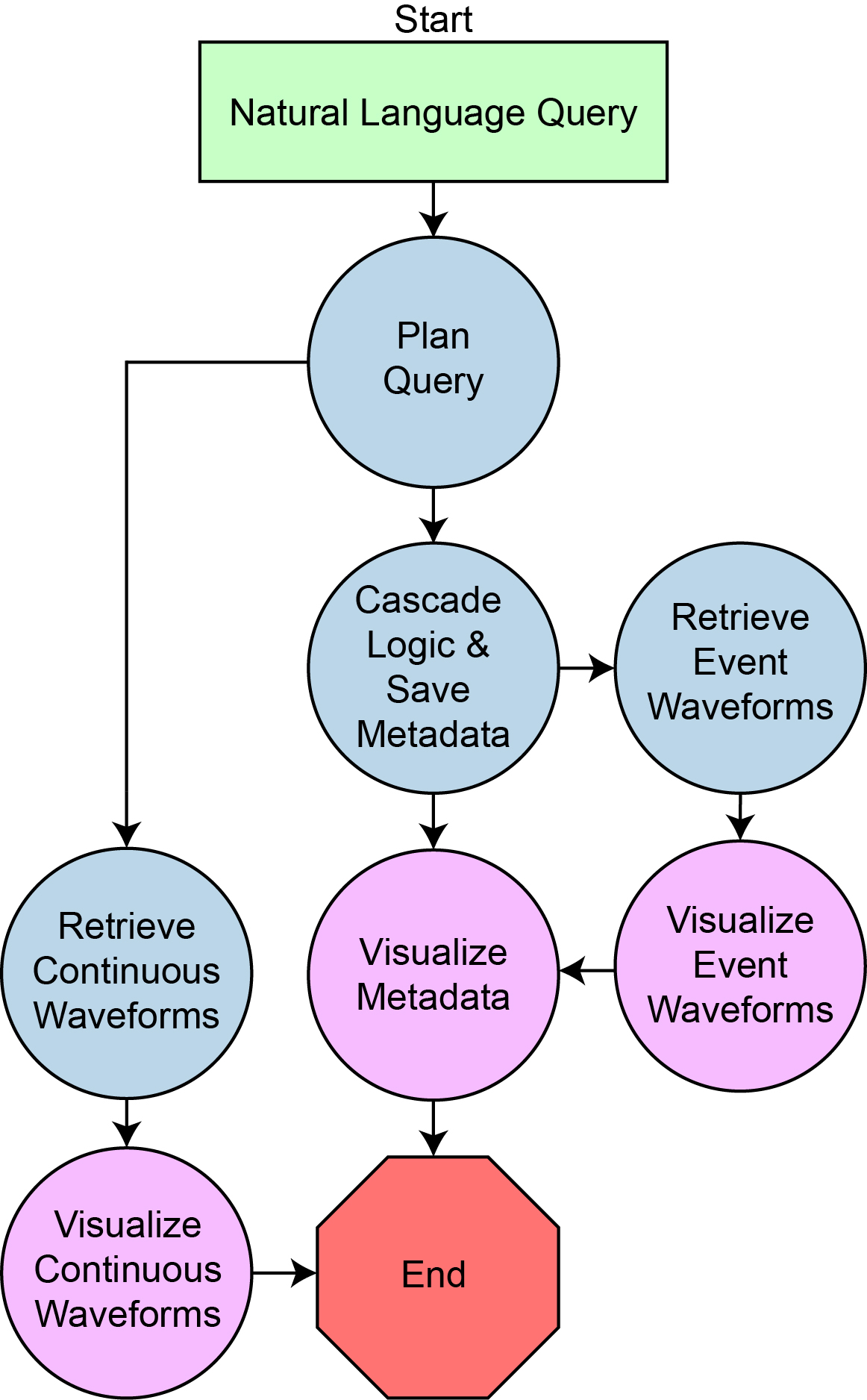}
    \caption{TREMORS LangGraph nodal structure. The graph represents a simplified representation of the possible decision tree pathways. The path is determined by LLM reasoning from the natural language query that the user prompts the system with.}
    \label{fig:1}
\end{figure}

\subsection{Example Workflows}
We now explore two examples associated with the two primary branches of the LangGraph decision tree (Figure \ref{fig:1}).
The agent execution is performed  on two primary operational modes, event-based and continuous waveforms acquisition, depending on the intent inferred from the prompt.
Below is a summary of the routing sequence across both operational modes.

\subsubsection{\textbf{Event-based Acquisition}}
\begin{figure}
    \centering
    \includegraphics[width=1.0\linewidth]{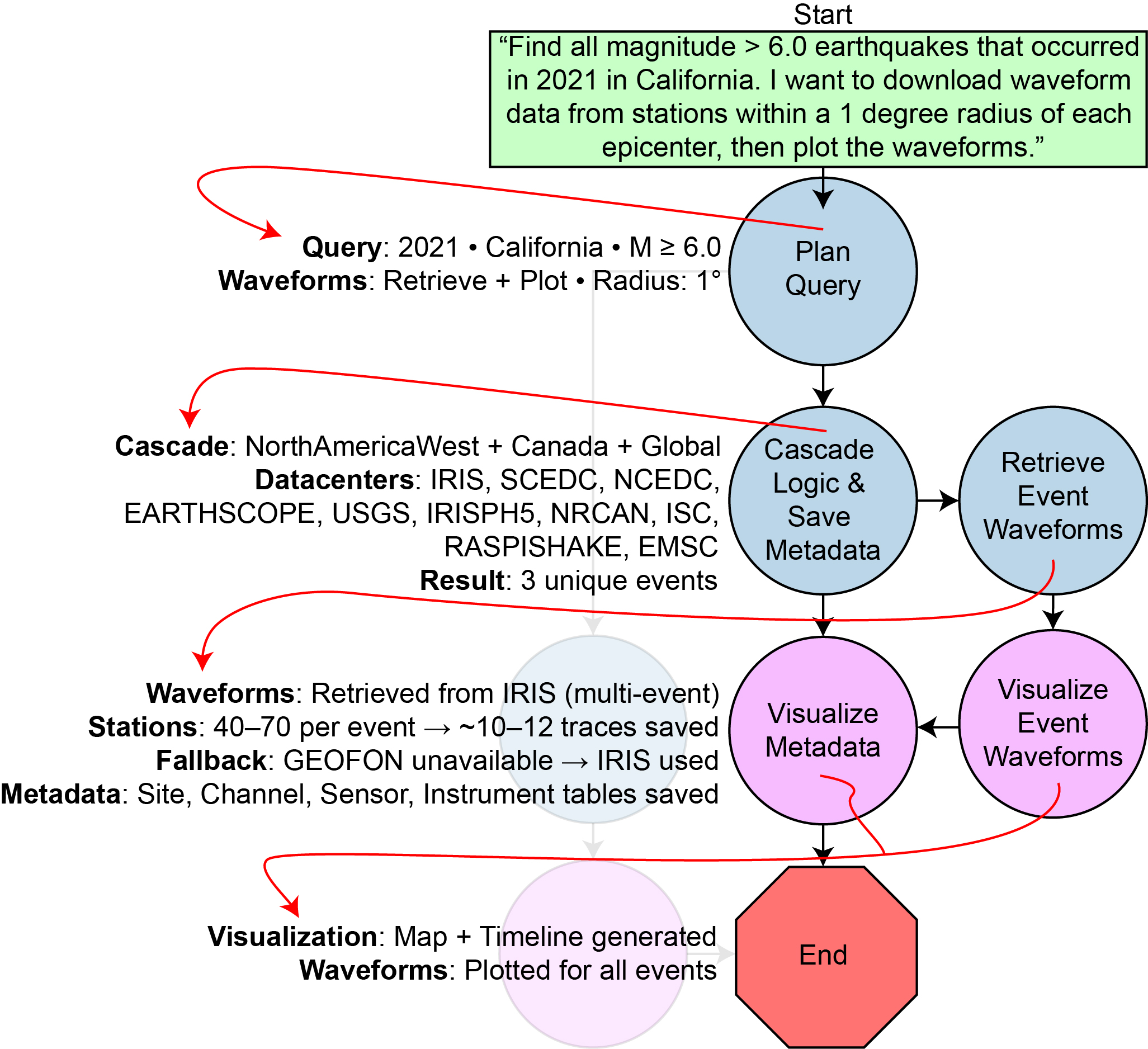}
    \caption{Example of an event-based waveform and plotting data acquisition result. The natural language query provides the context by which the LLM reasons. The query is first parsed and then determines the subsequent pathway. We show simplified example outputs as each step executes in the graph.}
    \label{fig:2}
\end{figure}

Figure \ref{fig:2} describes the graph execution pathway with simplified CLI outputs for the example prompt:
"\textit{Find all magnitude > 6.0 earthquakes that occured in 2021 in California. I want to download waveform data from stations within a 1 degree radius of each epicenter, then plot the waveforms}."

\begin{figure}
    \centering
    \includegraphics[width=1.0\linewidth]{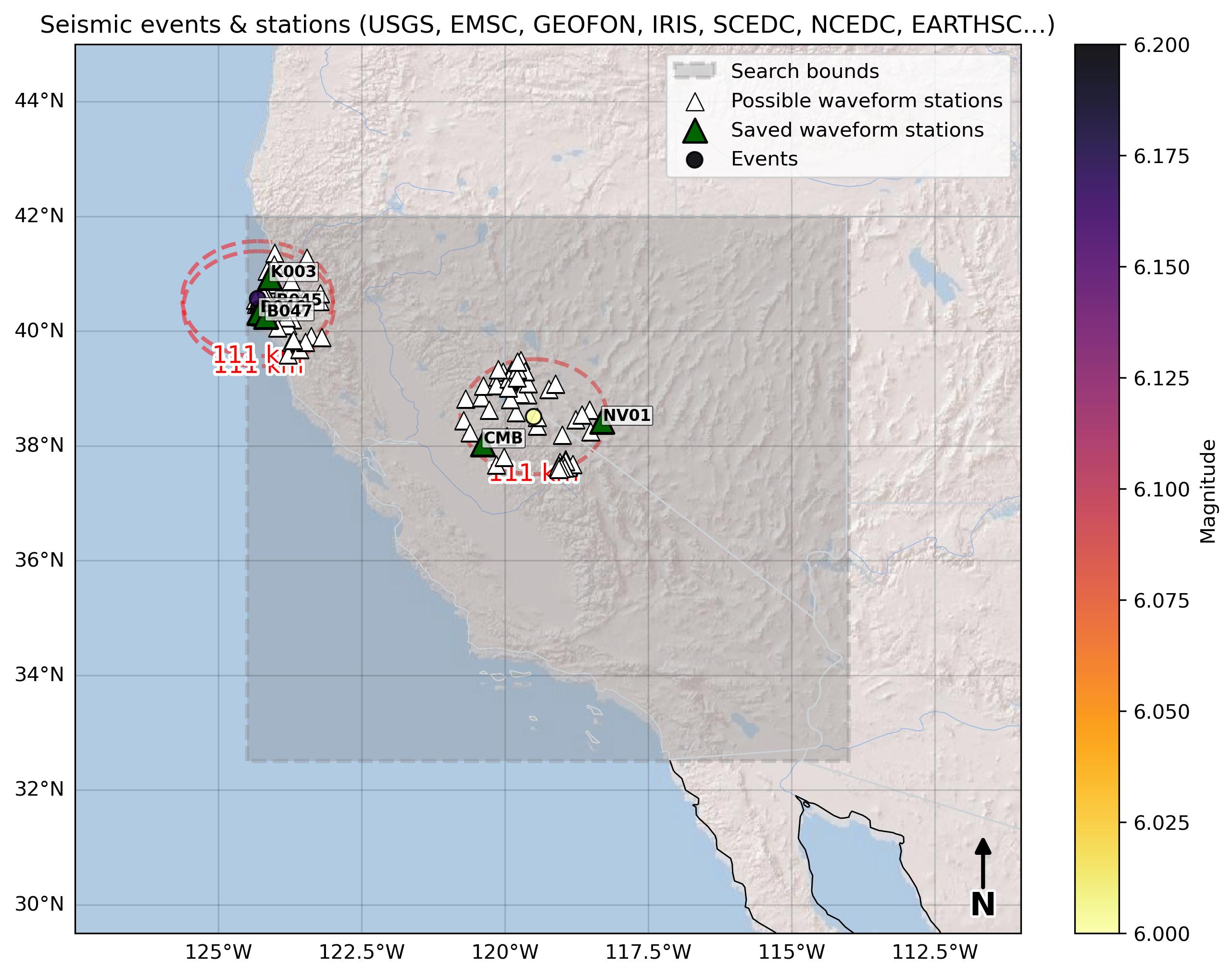}
    \caption{Example plot associated with the prompt in Figure \ref{fig:2}. Based on language in the prompt a bounding box over California is estimated (gray box). The three events are plotted and colored by magnitude (circles). Bounds on the possible stations around each epicenter are also plotted (red-dash). All possible stations that are found in the multi-datacenter acquisition are plotted (triangles). Waveforms are saved up to the default limit of 10 traces for each event (occasionally more if a station has multiple channels) and are plotted for the first associated stations until this limit is reached (green triangles). This default limit may be increased by the user.}
    \label{fig:3}
\end{figure}

Each node shown in Figure \ref{fig:2} represents a particular execution that occurs during the request.
\begin{enumerate}
    \item \textbf{Plan Query} (Entry Point): The natural language prompt is passed to a constrained LLM which extracts explicit search thresholds into a standardized JSON schema. From the prompt, it sets the min/max dates for 2021, builds a bounding box around California (reasoned by the LLM), sets the minimum magnitude, sets a search bounding radius of 1 degree, and explicitly flags both waveform retrieval and plotting as \texttt{True}.
    \item \textbf{Check Continuous Waveform Request} (Conditional Edge): The graph evaluates the extracted state realizing that \texttt{get\_continuous\_waveforms} is \texttt{False}, routing execution to the standard cascade node.
    \item \textbf{Cascade Logic \& Save Metadata}: The agent triggers an exhaustive, cascading search querying multiple predefined target datacenters based on the regional match. It adds the datacenters associated with our predefined `NorthAmericaWest' region that includes IRIS, SCEDC, NCEDC, EARTHSCOPE, USGS, IRISPH5. From the bounding box overlap with the regional perimeters it also includes Canada datacenter NRCAN. Always included are the global datacenters ISC, RASPISHAKE, and EMSC. The query searches for seismic events that match the geographic box and magnitude limits. Unique overlapping records are deduplicated. The composite ObsPy catalog objects are then passed through a metadata transformation method which parses the deeply nested FDSN objects and flattens them into a seismic database schema based on the NNSA Knowledge Base \citep{carr2002national} with slight modifications. This isolates attributes into relational tables (e.g., event, origin, origerr, netmag) that are saved using Parquet file formats for efficient downstream queries and relational joins.
    \item \textbf{Retrieve Event Waveforms} (Conditional Edge): The graph invokes waveform retrieval and recognizes waveform plotting is triggered. This node accesses the discovered earthquake metadata, searches for any station within the provided 1 degree epicenter radius, and retrieves matching MiniSEED waveforms. The retrieved physical trace stations are subsequently cataloged into an output Parquet file.
    \item \textbf{Visualize Waveform/Metadata} (Terminal Node): Since the prompt asks for plots, the framework calls the plotting node to produce a map view summary of the results. The map plots earthquake locations within the specified bounding box, overlays radius boundary lines, and displays station networks colored by waveform availability to distinguish those that successfully yielded trace data. These parameters can be adjusted by the user, with a default value of 10 traces (Figure \ref{fig:3}).
\end{enumerate}

\subsubsection{\textbf{Continuous Waveform Acquisition}}
When a user is interested in conditions regardless of distinct physical event signatures (i.e. ambient noise or explicit time-frames), they will instead invoke the continuous waveform pathway.
This method can find stations within a specified box, but may also explore a provided stations file, as seen with the example prompt:

"\textit{Get continuous waveforms using the stations file teststations.txt. Do not use directory dates, use directory stats, and download response.}"

The teststations.txt file is available in the TREMORS repository, but each row is a single station including network, station, location, channel, datacenter, time window (e.g., \textit{IU ANMO BH 00 IRIS 2016-01-01T00:00:00 2016-01-03T00:00:00
}). Node execution follows the following path:
\begin{enumerate}
    \item \textbf{Plan Query} (Entry Point): The natural language prompt is passed to the constrained LLM, which extracts an alternate configuration schema tailored for continuous data retrieval. Recognizing the explicit request for continuous waveforms, the agent sets `get continuous waveforms' as \texttt{True}, identifies the user-provided station list, configures any specified temporal constraints, and enables response metadata retrieval as requested.

    \item \textbf{Check Continuous Waveform Request} (Conditional Edge): The graph evaluates the extracted state. This acts as a critical control flag, overriding the standard event-based workflow. Execution is immediately rerouted to the continuous data branch ignoring all earthquake catalog queries.

    \item \textbf{Retrieve Continuous Waveforms}: In this execution path, earthquake event queries and spatial bounding boxes are ignored entirely. Instead, the node reads the provided text file list manifests and parses the specified networks, stations, and channels. The agent then directly interfaces with FDSN services to retrieve continuous waveform data streams based solely on the provided time window and station availability. Raw waveform segments are downloaded in bulk, and associated instrument response metadata is simultaneously retrieved and attached.

    \item \textbf{Plot Continuous Waveforms} (Terminal Node): After retrieval and caching of the continuous waveform data, the graph proceeds to a rudimentary plotting stage. The system processes the unsegmented time-series data, optionally applies instrument response correction (deconvolution), and generates visualizations of the ambient seismic signals.
\end{enumerate}

\section{Discussion}
As AI-driven tools and modeling capabilities continue to advance, the need for data systems that support scalable querying, efficient access, and seamless sharing across networks will become increasingly critical.
TREMORS represents an advancement in observational seismology for data acquisition and provides a blueprint for reproducible multi-step data procurement and analyses across multiple repositories.
The key contribution is the use of LLM-guided semantic parsing within a constrained execution graph to mediate between scientific intent and distributed seismic data infrastructure.
In this framework, we adopt a relational storage scheme in which metadata are stored in Parquet format, while seismic waveforms are stored in MiniSEED files (with daily segmentation for continuous data).
This design reflects a balance between efficient querying and compatibility with existing seismological standards, although it is not necessarily optimal.
More broadly, the question of best practices for seismic data storage, structure, and access remains open.
Thus, TREMORS should be viewed as retrieval utility and template of how seismological data workflows need to shift toward AI-ready infrastructure

As agentic workflows appear in many scientific disciplines, seismic analysis stands to benefit from increased automation and acceleration in this new era of scientific agents \citep{ren2025seismologymodelingagentsmart,liu2026tracemultiagentautonomousphysical,xi2025rise}.
This is fortuitous for seismology since it is uniquely positioned to benefit from agentic workflows due to its extensive archive of well-documented metadata and high-fidelity data.
While agents are not yet capable of replacing high-level subject matter experts or independently producing scientific discovery, TREMORS represents an important first step in assisting scientists with routine data retrieval tasks, enabling greater focus on inquiry and discovery.
Like most agents, TREMORS is highly effective at accelerating routine tasks and data retrieval.
However, the term \textit{agent} is currently used broadly, so we emphasize that our implementation is explicitly structured as a graph of states and transitions, rather than a simple linear prompt-response system.
In this design, predefined nodes (reasoning steps), edges (decision logic), and states (shared, evolving memory) form a controlled and transparent workflow.
This structure enables the encoding of iterative reasoning directly into the process providing explicit control over decision making.
It also facilitates integration of external tools, models, or post-processing that users may incorporate as additional graph nodes.
The architecture allows for reproducibility due to the traceable path of the decision graph - a significant importance to scientific workflows.

Despite its practical utility, TREMORS is subject to limitations.
While the chosen agentic framework reduces uncertainty associated with natural-language interpretation, it does not eliminate it.
The semantic parsing stage may fail for ambiguous, imprecise, or unusually phrased prompts.
Although execution through the graph is deterministic once parameters are extracted, the correctness of the resulting workflow still depends on the validity of the parsed schema and on assumptions embedded in default parameter choices.
For this reason, users benefit from understanding the expected parsing structure described in the Methods.
In addition, TREMORS depends on external FDSN services and ObsPy-compatible interfaces, making it sensitive to heterogeneous metadata standards, incomplete station information, service outages, and variability across datacenters.
Although this is mostly accounted for with our kbcore-style metadata transformation \citep{carr2002national} and station querying functionality, the system remains constrained by the behavior and completeness of the underlying services.
This limitation is important conceptually---the central challenge is not any single protocol, but the full problem of building retrieval systems that can operate robustly across evolving heterogeneous archive ecosystems.
Improvements will focus on validation across diverse use cases and prompting styles, while expanding support to additional protocols for non-FDSN compliant datacenters.

We view schema focused agent driven retrieval as a promising direction for the field.
If seismology is to fully leverage emerging AI capability it will benefit from data access workflows that are available across many repositories.
They must also satisfy the requirement of adaptability to future APIs rather than any single service interface.
To this end, TREMORS is released as an open source user-centered design.
It is not intended as a final solution to seismic data procurement, but a starting point demonstrating how agentic systems will broaden participation through reduced user burdens and support reproducible seismological research.

\section{Conclusions}
We present TREMORS (Text Referenced Event Mapping and Output Renderer for Seismographs), an agent-based framework designed to automate and accelerate seismic data retrieval from FDSN-compliant datacenters.
By combining semantic parsing with a deterministic execution graph, the system translates natural language queries into reproducible workflows for multi-datacenter seismic data acquisition.
The event-based and continuous waveform examples demonstrate the model's data acquisition across multiple datacenters highlighting its flexibility and practicability.
While the current state of the agent is focused on waveform and metadata acquisition, the graph structure is intentionally modular to incorporate additional post-processing steps, analysis, or modeling steps.
TREMORS illustrates how future seismological data systems should be organized around portable schemas, constrained workflow layers, and adaptable multi-repository access rather than around any single protocol or interface.
It also represents a step toward AI-ready seismological workflows that remain cumbersome for many scientists, thereby, lowering user burden and strengthening scientific data infrastructure and enabling more efficient data-driven discovery.

\section*{Data and Resources}
All materials presented including source code files are available online Hill, R.G. \citeyear{HillZenodoTREMORS}.

\section*{Declaration of Competing Interests}
The authors acknowledge that there are no conflicts of interest recorded.

\section*{Acknowledgments}
This material is based upon work supported by the U.S. Department of Energy, Office of Science, Office of Basic Energy Sciences, Geosciences program under Award No. LANLE3CB to support RGH, RAD, and CWJ in all aspects of this manuscript. Approved for unlimited release LA-UR-26-23557.

\bibliographystyle{plain}
\bibliography{tremor}

\end{document}